\documentstyle[lprocl,11pt]{article}

\def\Journal#1#2#3#4{{#1} {\bf #2}, #3 (#4)}

\def\PLB{{\em Phys. Lett.}  B}

\def\PRD{{\em Phys. Rev.} D}
\def\PR{\em Phys. Rev.}

\def\ra{\rightarrow}

\def\be{\begin{equation}}
\def\ee{\end{equation}}
\def\bea{\begin{eqnarray}}
\def\eea{\end{eqnarray}}

\begin{document}

\title{HIGHER ORDER RENORMALONS\footnote{Presented at 
VIIth Blois Workshop, Seoul, Korea, June 10-14, 1997.}}
\author{ TAEKOON LEE }

\address{ Department of Physics, Purdue University, \\ West Lafayette,
 IN 47907}

\maketitle\abstracts{ A systematic method of summing the corrections
to the renormalon residue arising from higher order renormalons is
discussed.}

\section{Introduction}

It is well known that the weak coupling expansion in field theory
is not convergent but asymptotic. Its coefficients diverge factorially
in the order of perturbation. One source of the divergence is renormalon.
For example, when a Green's function such as the Adler function $D(\alpha)$ 
is expanded as
\be
D(\alpha)=\sum a_{n} \alpha^{n+1},
\label{1}
\ee
the infrared renormalon causes the coefficients to grow as
\be
a_{n} \ra K\, n!\, n^{-2 \beta_{1}/\beta_{0}^{2}}\, (-2/\beta_{0})^{-n},
\mbox{\hspace{.3in} for $ n \ra \infty$},
\label{2}\ee
where $ \beta_{0}, \beta_{1}$  are the first two coefficients of the $ \beta$
function. 

The constant $K$ is an all order quantity: it gets contribution not
only from the well-known renormalon diagram, a chain of one loop
bubbles, but also from infinitely many higher order renormalon
diagrams. The purpose of this talk is to show summing these
corrections is manageable, and that   $K$ can be expressed
in a calculable, convergent sequence.  
 
\section{Renormalons in QED }
For our purpose it is convenient to consider the Adler function
defined by
\bea
D(\alpha(q^{2}))&=&\frac{\partial}{\partial q^{2}} \Pi(q^{2}) \nonumber \\
                &=& \mbox{const.} + \sum_{n=0}^{\infty}
                             a_{n}(\mu^{2}/q^{2}) \alpha(\mu)^{n+1}
\eea
where $\Pi$ is the vacuum polarization function of the
electromagnetic current and $\mu$ is the renormalization scale.

The large order behavior in Eq.~\ref{2}  arises from an exchange of 
the all-order  Gell-Mann-Low (GL) effective charge \cite{lau}
defined by \cite{gl}
\begin{equation}
{\mathbf{a}}(\alpha(\mu),t) = \frac{\alpha(\mu)}{1 + 
\Pi(\alpha(\mu),t)}
\label{4} \end{equation}
where $t=k^{2}/\mu^{2}$.
Therefore, as far as we are concerned with the infrared renormalon 
associated with the asymptotic behavior, we can write the Adler
function as
\begin{equation}
D(\alpha(\mu),\mu^{2}/q^{2}) = \int_{0}\, f( k^{2} )\,
{\mathbf{a}} (k^{2})\, d\, k^{2}
\label{5} \end{equation}
where $ {\mathbf{a}} ( k^{2} )$ denotes  the GL effective charge and
\begin{equation}
f (k^{2})  =   \frac{ -e_{f}^{2} k^{2}}{
8 \pi^{3} q^{4}} \hspace{.5in}  \mbox{ for $k^{2} \rightarrow 0$}
\label{6} \end{equation}
with $e_{f}$ denoting  the  electric charge.

The divergence in Eq.~\ref{2} causes a singularity in Borel plane. The Borel
transform  of $D(\alpha(\mu),\mu^{2}/q^{2})$ is defined by
\be
D(\alpha(\mu),\mu^{2}/q^{2})=\int_{0}^{\infty} \exp \left(-\frac{b}{
\alpha(\mu)}\right)
\widetilde{D}(b)\,d\,b.
\label{35} \ee
With the perturbative series 
in Eq.~\ref{1}  the Borel transform becomes
\be
 \widetilde{D}(b)=\sum_{n=0}^{\infty} \frac{a_{n}}{n!} b^{n}.
\label{36} \ee
Substitution of the asymptotic form in Eq.~\ref{2} into Eq.~\ref{36} gives
\be
\widetilde{D}(b) \ra \frac{K (-2 \beta_{1}/\beta_{0}^{2})!}{
\left(1+\beta_{0} b/2\right)^{1-2 
\beta_{1}/\beta_{0}^{2}}}
\ee
in the neighborhood of the singularity at $b = -2/\beta_{0}$. Thus,
to calculate $K$ we  need to evaluate the residue of the
singularity.

Let us now consider the renormalization group equation for the
GL effective charge:
\be
\left( \mu^{2} \frac{\partial}{\partial \mu^{2}} +
\beta(\alpha)\frac{\partial}{\partial\alpha} \right) {\mathbf{a}}(
\alpha(\mu),t)=0.
 \ee
Solving the equation we may write the GL effective charge as
\be
 {\mathbf{a}}(k^{2})= \frac{1}{\frac{1}{A(k^{2})} + C({\mathbf{a}}(k^{2}))}
\label{38} \ee
where $ C({\mathbf{a}})$ is a scheme independent function.
The effective coupling $A$ is defined by
\be
A(\alpha(\mu),t)=\frac{1}{ - \beta_{0} \left( \ln t +
 \int^{\alpha(\mu)} \frac{1}{ \beta(\alpha)}\,d\,\alpha -\frac{
p_{1}}{\beta_{0}}\right)},
 \ee
where $ p_{1}$ is the constant term in the one loop vacuum
polarization function.
The function $C({\mathbf{a}}(k^{2}))$  may be expanded 
in an asymptotic series  as
\be
C( {\mathbf{a}}(k^{2}) ) = c_{1} \ln ( {\mathbf{a}}(k^{2}) ) + 
\sum_{2}^{\infty} c_{i}
\,[ {\mathbf{a}}(k^{2})]^{i-1}.
 \ee
The  $ c_{i}$ can be easily determined in terms of the coefficients
of the $\beta$ function and the vacuum polarization function.

Now define an $N$ th order GL effective charge:
\be
{\mathbf{a}}^{\left(N\right)}(k^{2}) =   \frac{1}{\frac{1}{A(k^{2})} + 
c_{1} \ln ( {\mathbf{a}}^{\left(N\right)}(k^{2}) ) + 
\sum_{2}^{N} c_{i} {\mathbf{a}}^{\left(N\right)}(k^{2})^{i-1}}.
\label{14} \ee
and introduce a modified Borel transform
\be
{\mathbf{a}}^{\left(N\right)}( \alpha(\mu),t)=\int_{0}^{\infty} 
\exp\left(-\frac{b}{A(t)}\right) 
\widetilde{{\mathbf{a}}}^{\left(N\right)}(b) \,d\,b.
\label{48} \ee
Substituting Eq.~\ref{48}  into Eq.~\ref{5}, we can 
write  $D^{\left(N\right)}(\alpha(\mu),\mu^{2}/q^{2})$,
 which is defined by replacing ${\mathbf{a}}(k^{2})$ in Eq.~\ref{5} 
 with ${\mathbf{a}}^{\left(N\right)}(k^{2})$, as 
\be
D^{\left(N\right)}(\alpha(\mu),\mu^{2}/q^{2}) =  \int \exp\left[ b
 \beta_{0} \int^{\alpha(\mu)}\frac{d\, \alpha}{\beta(\alpha)}\right]
\left\{ e^{-b p_{1}} \widetilde{f}(b) \widetilde{{\mathbf{a}}}^{\left(N\right)}
(b)\right\} d\,b,
\label{49} \ee
where 
\be
\widetilde{f}(b) = \int_{0} f(t) \exp\left( b \beta_{0} \ln t\right) \,d\,t.
\label{50} \ee
The first IR   renormalon singularity
  arises from the IR  divergence in the integral  in Eq.~\ref{50}.
 Substituting Eq.~\ref{6}  into Eq.~\ref{50}
\bea
\widetilde{f}(b) &=& \int_{0}^{M} f (t) e^{ b \beta_{0} \ln t} \,d\,t
 \nonumber\\
&=&-\frac{ e_{f}^{2} \mu^{4}}{8 \pi^{3} q^{4}} \int_{0} ^{M} t 
 e^{ b \beta_{0} \ln t}
\,d\,t \nonumber \\
&=&-\frac{ e_{f}^{2} \mu^{4}}{8 \pi^{3} q^{4}}\frac{1}{2 + b
 \beta_{0}} \left(
1 + ( 2 + b \beta_{0}) \ln M + \cdots  \right),
\label{51} \eea
where $M$  is an arbitrary UV  cutoff.
Notice that the leading renormalon singularity is cutoff 
independent.

In the renormalization scheme in which the $\beta$ function is 
given by
\be
\beta(\alpha)=\frac{\beta_{0}  \alpha^{2}}{1-\lambda \alpha}
\label{522} \ee
with $\lambda=\beta_{1}/\beta_{0}$,
the Eq.~\ref{49} becomes the modified Borel transform introduced by
Brown, Yaffe, and Zhai \cite{zhai}.
Then using  the relation between the ordinary Borel transform 
 and  the modified one, we have the ordinary Borel 
transform
\be
\widetilde{D}^{\left(N\right)}(b) \ra
-\frac{ e_{f}^{2} \mu^{4}}{16 \pi^{3} q^{4}} e^{- b_{0} p_{1}} 
 e^{-\lambda b_{0} \ln b_{0}} \widetilde{{\mathbf{a}}
}^{\left(N\right)}(b_{0})
 \frac{(-2\lambda /\beta_{0})!}{(1 +\frac{1}{2} b 
\beta_{0})^{1-2\lambda/\beta_{0}}}
\label{58} \ee
where $b_{0}= -2/\beta_{0}$.

Using Eq.~\ref{14} the Borel transform of the effective charge can be 
obtained without difficulty. The final expression is given by \cite{tlee}:
\bea
b^{bc_{1}}\widetilde{{\mathbf{a}}}^{\left(N\right)}(b)&=& \frac{b^{b c_{1}}}
{2 \pi i} \int e^{ b y}\, \sum_{k=0}^{\infty}
\frac{(-b)^{k}}{k!}\, \sum_{l=k}^{k (N-1)} \frac{h_{Nkl}}{y^{l}}\, \sum
_{i=0}^{N} \bar{c}_{i} y^{ b c_{1} - i-1} \,d\,y \nonumber \\
&=& \sum_{k=0}^{\infty} \,\sum_{l=k}^{k (N-1)} \,\sum_{i=0}^{N} \frac{
(-1)^{k}  h_{Nkl} \bar{c}_{i}}{
k!\, \Gamma(l+i+1-b c_{1})} b^{ k + l+i }
\label{72} \eea
where
\bea
\bar{c}_{i} =\left\{ \begin{array}{cc}  1 & \mbox{for $ i=0$} \\
c_{1} & \mbox{for $i=1$} \\
(i-1)\, c_{i} & \mbox{ for $ i \ge 2$},
\end{array}
\right.
\label{67} \eea
and
\be
h_{Nkl}= k! \sum_{\left\{n_{i}\right\}} \frac{
\prod_{i=1}^{N-1} c_{i+1}^{n_{i}}}{
\prod_{i=1}^{N-1} n_{i}!}
\label{69} \ee
with the set $\{n_{i}\}$ of nonnegative integers  satisfying
\be
\sum_{i=1}^{N-1}  n_{i} \cdot i  = l, \hspace{.5in} 
\sum_{i=1}^{N-1} n_{i}= k.
\label{70} \ee

To find the  renormalon residue of $\widetilde{D}^{\left(N\right)}(b)$,
we have to evaluate
$b^{ b c_{1}}\widetilde{{\mathbf{a}}}^{\left(N\right)}(b)$ at
the first IR renormalon position, $b_{0}=-2/\beta_{0}$.
If we directly
substitute $b$ in Eq.~\ref{72} with $b_{0}$, 
the resulting large order behavior does not have a  finite limit for
 $N \rightarrow \infty$ \cite{fs}. The reason for this is that  
 $\widetilde{{\mathbf{a}}}(b) $ is singular at the  UV and IR renormalon
positions, and its radius of convergence when it is expanded as in
 Eq.~\ref{72} 
is given by the position at $b=1/\beta_{0}$  of the first UV renormalon, 
which is the closest renormalon to the origin in  the Borel plane.
 Therefore we cannot
 substitute $b$ with $b_{0}$ in Eq.~\ref{72}
to correctly  evaluate the 
Borel transform  at the first
IR renormalon.

This problem can be avoided by introducing an analytic transform of the
Borel plane so that the closest renormalon to the origin in the
new complex plane  is the first IR renormalon~\cite{mueller3}.
Because the singularity of  $\widetilde{{\mathbf{a}}}(b) $ at the IR 
renormalon is
such that it is finite but has a  divergent derivative \cite{grunberg1},
we can then express the residue as a convergent series.

For this purpose,
 we can take any analytic transform that puts the IR renormalon
as the closest singularity to the origin. Here we  take:
\be
z= \frac{\beta_{0} b}{1-\beta_{0} b}.
\label{255}\ee
In the $z-\mbox{plane}$, the closest singularity to the origin is the
first IR renormalon at
\be
z_{0}=-\frac{2}{3},
\ee
and all the UV renormalons are pushed beyond $z=-1$ on the real axis.

To find  $ b^{bc_{1}}
\widetilde{{\mathbf{a}}}^{\left(N\right)}(b)$ at the first IR renormalon,
we have to substitute $b$ in Eq.~\ref{72}  with the inverse of Eq.~\ref{255}
 and expand it  in
Taylor series at $z=0$ to order $N$, and  evaluate it at
$z=z_{0}$.
Then the Borel transform  of GL effective charge at the first IR renormalon
is given by
\be
 \kappa_{N}=\left. b^{bc_{1}} \widetilde{{\mathbf{a}}}^{\left(N\right)}(b)
 \right|_{b=b_{0}} = \sum_{M=0}^{N} q_{M}\, z_{0}^{M} ,
\ee
where it is straightforward to express $q_{M}$ in terms of $c_{i}$.

Finally, the coefficient $K$ in Eq.~\ref{2} is given by
\be
K= \lim_{N\ra \infty} K_{N}
\ee
where
\be
K_{N}  = -\frac{e_{f}^{2} \mu^{4}}{16 \pi^{3} q^{4}} e^{
 - b_{0} p_{1}}  \kappa_{N}.
\label{76} \ee
The first four elements of the sequence is given in Table 1 for 
different number of fermion flavors.
\begin{table}
\caption{The first four elements of the sequence for the first IR renormalon
residue in QED}
\vspace{0.4cm}
\centering
\begin{tabular}{|c|c|c|c|c|c|c|} \hline
& $N_{f}=1$ & $N_{f}=2$ & $N_{f}=3$ & $N_{f}=4$ & $N_{f}=5$ &$N_{f}=100$ \\ \hline
$\kappa_{0}$ &1  &1 &1 &1 &1&1 \\ \hline
$\kappa_{1}$ &1.63  &1.32 &1.21 &1.16 &1.13&1.00 \\ \hline
$\kappa_{2}$ &0.71  &1.31 &1.31 &1.27 &1.24&1.02 \\ \hline
$\kappa_{3}$ &-1.53 &1.25 &1.41 &1.39 &1.34&1.02 \\ \hline
\end{tabular}
\end{table}

\section{Renormalons in QCD}
In QCD,  there is unfortunately 
no satisfactory definition of renormalization 
scheme and scale invariant effective charge that may be used in the
diagrammatic study of renormalon. 
However, if we are only interested in the calculation of the 
residue, the definition
of the effective charge is not required. Indeed the calculation 
 is cunningly simple; it only requires the strength of the 
renormalon singularity and the perturbative calculation of $D(\alpha)$.
For this reason, the method described in the following can be 
equally well applied to the calculation of the UV renormalon 
residue.

Consider the Borel transform of the current correlation function in QCD.
The renormalon singularity of   $\widetilde{D}(b)$ in QCD
\be
 \widetilde{D}(b) \approx \frac{\widehat{D}}{ (1- b/b_{0})^{ 1 + 
\lambda b_{0}}}
\label{t1}
\ee
gives the large order behavior
\be
a_{n} \approx \frac{\widehat{D}}{(\lambda\, b_{0} )!} n! n^{\lambda\, b_{0}}
\, b_{0}^{-n}.
\ee

To calculate the residue $\widehat{D}$, consider a function
\be 
R(b) = \widetilde{D}(b)\, (1-b/b_{0})^{ 1 + \lambda b_{0}}.
\ee
Then because of Eq.~\ref{t1}, we have
\be
\widehat{D} = R(b_{0}).
\ee
To avoid the first UV renormalon,  we introduce  a new variable $z$, as we
did in QED, 
which is defined by
\be
z= -\frac{\beta_{0} b}{1-\beta_{0} b}
\ee
with its inverse
\be
b= \frac{-1}{\beta_{0}} \left( \frac{z}{1-z}\right).
\ee
In the $z-$plane, the IR renormalon at
\be 
z_{0}=\frac{2}{3}
\ee
is the closest singularity to the origin, and so the radius of convergence
of the Taylor series of  $\widetilde{D}(b(z))$ at $z=0$
is given by the first IR renormalon. 

Now $ \widehat{D}$
can be expressed in a convergent series form
\bea
\widehat{D} &=&\left. \widetilde{D}(b)\, (1-b/b_{0})^{ 1 + 
\lambda b_{0}}\right|_{b=b_{0}} \nonumber \\
&=& \left.\left(\sum_{n=0}^{\infty} \frac{a_{n}}{n!} \left[b(z)\right]^{n}
\right) (1-b(z)/b_{0})^{ 1 + \lambda b_{0}}\right|_{z=z_{0}} \nonumber \\
&=& \sum_{n=0}^{\infty} r_{n} z_{0}^{n},
\eea
where it is straightforward to find $r_{n}$ in terms of the 
perturbative coefficients $a_{n}$. Note that the series is 
convergent even if $R(b(z))$ is not analytic at $z=z_{0}$, because then
the radius of convergence of the series is given by $z=z_{0}$, and 
$R(b(z_{0}))$ is finite.

Using the  perturbative calculation of the current correlation
function, and $D(\alpha)$, to three loop \cite{zhai,kataev}, we have
\be
R(b(z))= \frac{3 \sum_{f} Q_{f}^{2}}{ 16\, \pi^3} \left[
1.333-0.748\,z -0.311 \,z^{2} +O(z^{3})\right]
\ee
for $N_{f}=3$. This is in the renormalization scheme in which the 
one-loop renormalization point is same as that of $\overline{MS}$ scheme,
and  the $\beta$
function is given in the form in Eq.~\ref{522}.
Evaluating this series at the renormalon position at $z=z_{0}$
we have
\bea
K_{1}&=&\frac{1.333}{(\lambda b_{0})!}= 0.946 \nonumber \\
K_{2}&=&\frac{(1.323-0.748\,z_{0})}{(\lambda b_{0})!}=0.592  \nonumber \\
K_{3}&=&\frac{(1.323-0.748\,z_{0}-0.311 \,z_{0}^{2})}
{(\lambda b_{0})!}= 0.494  \nonumber \\
\eea
For  several other flavor numbers we give $K_{n}$   in Table 2.
\begin{table}
\caption{The first three elements of the sequence for the large order behavior
in QCD}
\vspace{.4cm}
\centering
\begin{tabular}{|c|c|c|c|c|c|} \hline
& $N_{f}=1$ & $N_{f}=2$ & $N_{f}=3$ & $N_{f}=4$ & $N_{f}=5$ \\ \hline
$K_{1}$ &.881  &.904 &.946 &1.018 &1.132  \\ \hline
$K_{2}$ &.521  &.546 &.592 &.674 &.813  \\ \hline
$K_{3}$ &.592 & .549 &.494&.411 &.307 \\ \hline
\end{tabular}
\end{table}

\end{document}